%% file: Template.tex
\documentclass{article}
\usepackage{spconf,amsmath,graphicx,hyperref}
\input{preamble}

\title{ImpactMat: Continuous Material Estimation \\for Inverse Impact Sound Rendering}

\name{
  Hyebin Cho$^{1,*}$ \quad
  Bumsoo Kim$^{2,*}$ \quad
  Joon Son Chung$^{1}$
  \thanks{$^{*}$These authors contributed equally to this work.}
}
\address{
$^{1}$Korea Advanced Institute of Science and Technology, South Korea \\
$^{2}$Chung-Ang University, South Korea 
}

\begin{document}

\maketitle

\begin{abstract}
Impact sound rendering synthesizes the sound produced when a 3D object is struck, but practical renderers often rely on fixed material presets such as wood, plastic, or steel. These presets limit the range of impact sounds a renderer can express, while manually adjusting the underlying material parameters remains difficult without expertise in material acoustics. We therefore study inverse impact sound rendering: predicting material parameters from a reference impact sound so that a simulator can recreate a similar material response. To support this task, we introduce \textbf{ImpactMat}, a dataset and benchmark of single and blended material impact sounds paired with ground-truth material parameters. We further propose a feed-forward model that predicts these parameters from one or more recordings, using blended materials to learn smooth transitions between material types.
Experiments show that our method outperforms competitive baselines and enables re-rendering from real recordings without manual parameter tuning.
The project page is available \href{https://material-from-impact.github.io/material-from-impact/}{here}.
\end{abstract}

\input{sections/intro2}
\input{sections/method}
\input{sections/experiments}

\section{Conclusion}
We addressed inverse impact sound rendering by estimating material parameters from reference recordings to reproduce similar sounds on virtual objects. We introduced ImpactMat and, to our knowledge, the first dataset and feed-forward framework specifically designed for continuous material-parameter prediction in this setting, without per-example optimization. Experiments demonstrate accurate, robust estimation and perceptually plausible re-rendering from real recordings. Our framework connects recorded impacts to reusable material parameters for controllable physics-based rendering. Future work with more diverse real recordings and contact conditions could further reduce the synthetic-to-real domain gap.

\clearpage
\bibliographystyle{IEEEbib}
\bibliography{shortstrings,refs}

\end{document}

%% file: preamble.tex
\usepackage{spconf,amsmath,graphicx,hyperref}

\usepackage[utf8]{inputenc} 
\usepackage[T1]{fontenc}    
\usepackage{url}            
\usepackage{booktabs}       
\usepackage{amsfonts}       
\usepackage{nicefrac}       
\usepackage{microtype}      
\usepackage{xcolor}         
\usepackage{nicematrix}

\usepackage[accsupp]{axessibility}  

\usepackage{cite}
\usepackage{caption}
\usepackage{subcaption}
\usepackage{graphicx}
\usepackage{hyperref}
\usepackage{cleveref}

\usepackage{lipsum}
\usepackage{wrapfig}
\usepackage{array}
\usepackage{xcolor}
\usepackage{makecell}
\usepackage{enumitem}

\usepackage{algorithm}
\usepackage{algpseudocode}
\floatname{algorithm}{Algorithm}
\algrenewcommand\algorithmicrequire{\textbf{Require:}}
\algrenewcommand\algorithmicensure{\textbf{Ensure:}}

\usepackage{orcidlink}
\usepackage[dvipsnames]{xcolor}
\usepackage{booktabs}

\usepackage{pifont} 

\usepackage{colortbl}
\usepackage{multirow}
\usepackage{multicol}
\usepackage{listings} 
\usepackage{titletoc} 
\usepackage{fancyvrb}
\usepackage{tcolorbox} 
\usepackage[accsupp]{axessibility}  
\usepackage[subtle]{savetrees} 

\definecolor{lightgray}{rgb}{0.83, 0.83, 0.83}
\definecolor{Gray}{gray}{0.6}
\definecolor{aliceblue}{rgb}{0.94, 0.97, 1.0}
\definecolor{mistyrose}{rgb}{1.0, 0.89, 0.88}
\definecolor{backcolour}{rgb}{0.95,0.95,0.92}

\definecolor{lightgray}{rgb}{0.83, 0.83, 0.83}
\definecolor{Gray}{gray}{0.6}
\definecolor{aliceblue}{rgb}{0.94, 0.97, 1.0}
\definecolor{mistyrose}{rgb}{1.0, 0.89, 0.88}
\definecolor{backcolour}{rgb}{0.95,0.95,0.92}

\newcommand{\appendixref}[2]{%
  \if\sepappendix1%
    #1
  \else%
    #2
  \fi%
}

\usepackage{cleveref}

\crefname{equation}{Eq.}{Eqs.}
\Crefname{equation}{Equation}{Equations}

\crefname{figure}{Fig.}{Figs.}
\Crefname{figure}{Figure}{Figures}

\crefname{table}{Tab.}{Tabs.}
\Crefname{table}{Table}{Tables}

\crefname{section}{Sec.}{Secs.}
\Crefname{section}{Section}{Sections}

\crefname{algorithm}{Alg.}{Algs.}
\Crefname{algorithm}{Algorithm}{Algorithms}

\usepackage{etoolbox}
\AtBeginDocument{%
  \setlength{\abovedisplayskip}{3pt}%
  \setlength{\belowdisplayskip}{3pt}%
  \setlength{\abovedisplayshortskip}{2pt}%
  \setlength{\belowdisplayshortskip}{2pt}%
}

%% file: sections/intro2.tex
\section{Introduction}

Impact sound rendering synthesizes the sound a virtual object makes when struck. It provides auditory feedback for physical interactions in applications such as robotic manipulation, video games, and virtual or augmented reality~\cite{owens2016visually,10.1007/978-3-030-01267-0_34,chung2024t,yuan2024retrieval,ahn2023novel,fan2026physics}. Physics-based renderers generate these sounds from an object's geometry, material properties, and impact conditions~\cite{6788060,TraerJames2018Hiof,o2002synthesizing,ren2013example,jin2020deep,jin2022neuralsound,diaz2023rigid}. The material is especially important because it shapes how the object vibrates and how quickly the sound decays. Prior work shows that small changes to material parameters can produce perceptually distinct sounds for the same geometry~\cite{van2001foleyautomatic,giordano2006material}.

\begin{figure}[t]
    \centering    \includegraphics[width=1\linewidth]{
    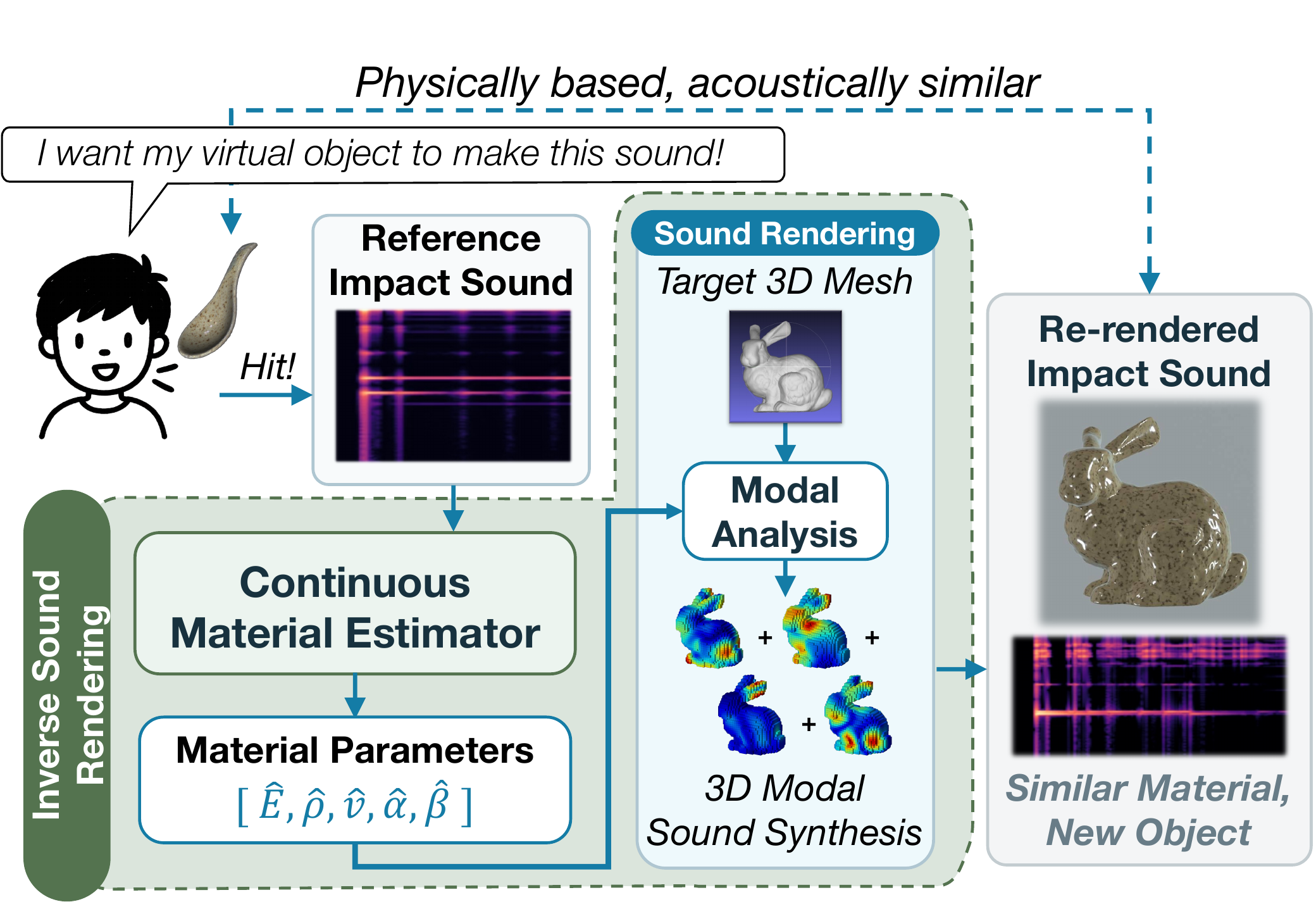}
    \caption{\textbf{Inverse impact sound rendering from reference audio.} Given a reference impact sound from an object with unknown material, our continuous material estimator predicts five renderer-compatible parameters, which are then used with a target mesh for physics-based sound synthesis. This enables re-rendering with a similar material response on a new object.}
    \label{fig:teaser}
    \vspace{-1em}
\end{figure}

\begin{figure*}[!t]
    \centering
    \includegraphics[width=\textwidth]{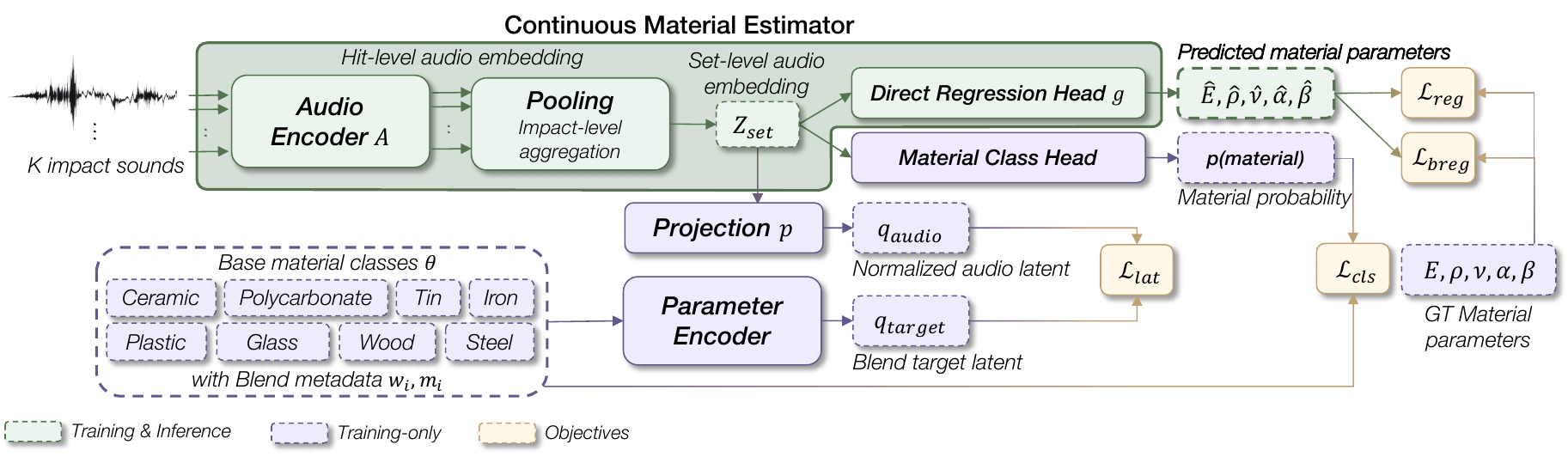}
    \vspace{-7mm}
    \caption{\textbf{Overview of the proposed continuous material estimator.} Multiple impact recordings are encoded, pooled as an unordered set, and mapped to continuous material parameters. Auxiliary classification and blend-consistency losses guide training, while inference uses the direct regression head.}
    \label{fig:method}
    \vspace{-5mm}
\end{figure*}

In practical rendering workflows, users often choose fixed material presets such as wood, plastic, or steel. These presets are convenient, but offer limited control over variation within a material class or intermediate material responses~\cite{5444993}. Moving beyond these presets requires manual tuning of material parameters, which demands expertise in material acoustics. Providing a reference recording of the desired sound may therefore be easier than specifying the parameters needed to reproduce it.

Inverse impact sound rendering addresses this difficulty by using a reference sound to infer the parameters needed by a renderer~\cite{zhang2017shape,diffimpact,jin2024diffsound}. In this setting, the goal is not only to recognize a material name, but to recover values that can synthesize a similar acoustic response. Differentiable rendering approaches can estimate such parameters by iteratively optimizing a renderer to match an input recording~\cite{zhang2017shape,jin2024diffsound}. However, this optimization must be repeated for each input, adding computational cost whenever a new recording is provided.

We focus on feed-forward continuous material-parameter estimation for inverse impact sound rendering (\cref{fig:teaser}). Given reference recordings, our model predicts renderer-compatible material parameters that can be used for controllable sound re-rendering, without manual tuning or per-recording optimization.

Our contributions are threefold. First, we formulate inverse impact sound rendering as continuous material-parameter estimation from reference audio. Second, we introduce \textbf{ImpactMat}, a dataset and benchmark of single- and blended-material impact sounds paired with continuous material parameters. Third, we propose a feed-forward framework that predicts these parameters from one or more recordings using blend supervision and set aggregation. Together, these components provide a benchmark and model for converting recorded impact audio into renderer-ready parameter estimates.

%% file: sections/method.tex
\section{Method}
\subsection{Dataset Generation}


We construct \textbf{ImpactMat} as a synthetic impact-sound dataset for inverse material estimation. We use 1,100 objects from ObjectFolder~\cite{gao2023objectfolder} and synthesize impact sounds with finite-element modal simulation, following~\cite{jin2022neuralsound}. For each object, we generate sounds for eight base materials: \textit{Ceramic}, \textit{Wood}, \textit{Plastic},
\textit{Polycarbonate}, \textit{Steel}, \textit{Tin},
\textit{Glass}, and \textit{Iron}. We additionally create two-material blends at ratios of 3:7, 5:5, and 7:3, excluding material pairs whose parameters are too similar to provide meaningful intermediate states. Each impact group contains multiple recordings with randomized hit point, force, and timing, so the same material can be observed under different contact conditions. Dataset statistics are summarized in \Cref{tab:dataset_stats}.

Each group is paired with a continuous material parameter vector $\theta=[\log E,\log \rho,\nu,\log \alpha,\log \beta]$, where $E$ is Young's modulus, $\rho$ is density, $\nu$ is Poisson's ratio, and $\alpha,\beta$ are Rayleigh damping coefficients. For blended materials, we linearly interpolate each physical parameter between
the two base materials using the specified mixing ratio,
before applying the log transformations. These blended examples cover intermediate material states and encourage a continuous acoustic-to-physical mapping rather than a lookup over fixed material labels.

\input{tables/dataset}
\vspace{-1mm}
\subsection{Architecture and Objective}

\input{tables/main_table}

Our model is designed around two requirements of inverse impact sound rendering: the estimate should be robust to contact-level variation, and the output should remain a continuous physical parameter vector rather than a material class. Since a single impact can vary with hit point, force, and timing, we treat multiple recordings from the same object-material condition as an unordered set. As illustrated in Fig.~\ref{fig:method}, each waveform is processed by a shared audio network $A$, consisting of a PANN CNN14 encoder~\cite{kong2020panns} followed by a projection MLP, to produce a 512-D hit embedding. We concatenate the mean and standard deviation of the hit embeddings and feed the result to a direct regression head $g$:
\begin{equation}
    \begin{aligned}
    z_k &= A(x_k), \quad k=1,\ldots,K,\\
    \mu_z &= \operatorname{mean}_{k}(z_k),
    \quad
    \sigma_z = \operatorname{std}_{k}(z_k),\\
    z_{\mathrm{set}} &= [\mu_z;\sigma_z],
    \quad
    \hat{\theta}=g(z_{\mathrm{set}}).
    \end{aligned}
\end{equation}

Target normalization statistics are computed on the training split, and predictions are denormalized for reporting and rendering. In our main model, final parameters come from this direct regression head.

We also attach a material classification head during training. For single-material sounds, the target is one-hot; for blend sounds, the target is given by the blend weights over the two base materials. This auxiliary head supplies weak category-level guidance, while inference remains a continuous regression problem rather than material classification.

Blend sounds discourage a lookup-like mapping to discrete material identities and provide continuous supervision in two ways. First, because blend samples have recomputed parameter targets, we apply an additional regression term on blend examples. Second, we use a latent blend consistency term that compares the audio latent of a blend with a target latent formed by mixing the two base-material embeddings using the blend weights. Let $p$ be the audio projection head, and let $q_i$ denote the normalized parameter embedding of base material $i$:
\begin{equation}
    \begin{aligned}
    q_{\mathrm{audio}} &= \operatorname{norm}(p(z_{\mathrm{set}})),\\
    q_{\mathrm{target}} &= \operatorname{norm}\!\left(\sum_i w_i q_i\right).
    \end{aligned}
\end{equation}
The latent loss is the cosine distance between $q_{\mathrm{audio}}$ and $q_{\mathrm{target}}$. The final objective is
\begin{equation}
    \mathcal{L}
    = \mathcal{L}_{\mathrm{reg}}
    + \lambda_{\mathrm{cls}}\mathcal{L}_{\mathrm{cls}}
    + \lambda_{\mathrm{breg}}\mathcal{L}_{\mathrm{breg}}
    + \lambda_{\mathrm{lat}}\mathcal{L}_{\mathrm{lat}} .
\end{equation}
Here, $\mathcal{L}_{\mathrm{reg}}$ is a target-weighted Smooth L1 loss over all training samples, $\mathcal{L}_{\mathrm{cls}}$ is soft-label cross entropy for the auxiliary material head, $\mathcal{L}_{\mathrm{breg}}$ is the same regression loss restricted to blend samples, and $\mathcal{L}_{\mathrm{lat}}=1-q_{\mathrm{audio}}^\top q_{\mathrm{target}}$. In the main configuration, $\lambda_{\mathrm{cls}}=0.25$, $\lambda_{\mathrm{breg}}=0.10$, $\lambda_{\mathrm{lat}}=0.05$, and the latent blend loss starts after a 30-epoch warm-up. At inference time, the same architecture accepts any evaluated $K$ by recomputing the set statistics.

%% file: tables/dataset.tex
\begin{table}[t]
\centering
\caption{\textbf{ImpactMat dataset and split statistics.} Single-material data are synthesized for each object using eight base materials. Blended-material data cover 24 selected material pairs at three mixing ratios, yielding 72 blend groups per object. Each impact group contains eight recordings.}
\label{tab:dataset_stats}
\vspace{-4pt}
\begin{tabular*}{0.9\columnwidth}{@{\extracolsep{\fill}}ccccc@{}}
\toprule
Split & Objects & Single & Blend & Total \\
\midrule
Train & 880 & 7,040 & 63,360 & 70,400 \\
Val   & 110 &   880 &  7,920 &  8,800 \\
Test  & 110 &   880 &  7,920 &  8,800 \\
\midrule
All   & 1,100 & 8,800 & 79,200 & 88,000 \\
\bottomrule
\vspace{-10mm}
\end{tabular*}
\end{table}

%% file: tables/main_table.tex
\begin{table*}[!t]
\centering
\caption{\textbf{Comparison of physical-parameter estimation methods.} Results on ImpactMat. $\rho$ and $E$ are reported in the log-space target scale, $\nu$ is linear, and $\alpha$ and $\beta$ in physical units. Avg.\ NMAE is computed over all five normalized target
dimensions and is omitted for DiffSound, which estimates
only $E$ and $\nu$ in our evaluation. Evaluated with \(K=1\), lower is better.} 
\label{tab:main_seen}
\vspace{-4pt}
\setlength{\tabcolsep}{2.5pt}
\begin{tabular}{lllcccccccc}
\toprule
 Method & Setting & Test subset & $\rho$ & $E$ & $\nu$ & $\alpha$ & $\beta$ & Avg.\ NMAE & Infer.\ time  & VRAM \\
\midrule
\multirow{2}{*}{DSP + Ridge Reg.} & \multirow{2}{*}{Conventional} & Single & 0.190 & 0.443 & 0.047 & 9.93 & $3.26{\times}10^{-7}$ & 1.405 & \multirow{2}{*}{16.1 (ms)} & \multirow{2}{*}{CPU} \\
& & Blend & 0.131 & 0.321 & 0.032 & 5.19 & $1.83{\times}10^{-7}$ & 0.961 &   \\
\addlinespace[2pt]

\multirow{2}{*}{Qwen2.5-Omni-7B \cite{xu2025qwen25omnitechnicalreport}} & \multirow{2}{*}{ICL} & Single & 0.338 & 0.614 & 0.071 & 13.38 & $1.61{\times}10^{-1}$ & 1.270 & \multirow{2}{*}{5.6 (s)} & \multirow{2}{*}{16.1 GB} \\
& & Blend & 0.257 & 0.495 & 0.067 & 8.23 & $1.73{\times}10^{-1}$ & 1.162 &   \\
\addlinespace[2pt]
\multirow{2}{*}{Audio Flamingo 3 \cite{ghosh2026audio}} & \multirow{2}{*}{ICL} & Single & 14.1 & 2.14 & 0.130 & 21.7 &  $1.8{\times}10^{-2}$ & 12.42 & \multirow{2}{*}{4.4 (s)} & \multirow{2}{*}{18.3 GB}\\
& & Blend & 13.4 & 2.33 & 0.123 & 11.3 & $4.4{\times}10^{-2}$ & 10.92 &   \\
\addlinespace[2pt]
\multirow{2}{*}{DiffSound \cite{jin2024diffsound}} & \multirow{2}{*}{Optimization} & Single & N/A & 	0.893 & 0.890 & N/A&  N/A & -- & \multirow{2}{*}{190 (ms)} & \multirow{2}{*}{2.0 GB}\\ 
& & Blend & N/A & 0.487 & 0.696 & N/A & N/A & -- &  \\ 
\addlinespace[2pt]
\rowcolor{aliceblue}
& & Single & 0.081 & 0.140 & 0.008 & 0.59 & $2.39{\times}10^{-8}$ & \textbf{0.078} &  & \\
\rowcolor{aliceblue}
\multirow{-2}{*}{\cellcolor{aliceblue}\textbf{Ours}} & \multirow{-2}{*}{\cellcolor{aliceblue}Feed-forward} & Blend & 0.119 & 0.206 & 0.012 & 1.08 & $1.32{\times}10^{-8}$ & \textbf{0.116} & \multirow{-2}{*}{6.1 (ms)} & \multirow{-2}{*}{1.33 GB}\\
\bottomrule
\vspace{-7mm}
\end{tabular}
\end{table*}

%% file: sections/experiments.tex
\section{Experiments}

\subsection{Experimental Setup}
\label{sec:experimental_setup}

We evaluate continuous material parameter estimation on test split of ImpactMat. 
Audio clips are resampled to 32 kHz and cropped to 2 s. We fine-tune the PANN CNN14 encoder for 200 epochs with AdamW, using $K=4$ impacts per training sample, and select the best validation checkpoint for evaluation. Analysis tables report normalized MAE averaged over the five targets.

The audio-language baselines use a reference set of 32 material configurations: eight base materials and 24 two-material blends with mixing weight $w=0.5$, each paired with its ground-truth parameters. Qwen2.5 receives these references as audio–parameter exemplars for in-context learning (ICL), whereas Audio Flamingo 3 receives a text-only parameter table due to its single-audio input constraint. 
DiffSound~\cite{jin2024diffsound} is initialized with the mean parameters of the eight base materials.

\vspace{-1mm}

\subsection{Physical Parameter Estimation}
\Cref{tab:main_seen} compares our model with hand-crafted feature regression, audio-language-model ICL, and optimization-based inverse rendering. Our model achieves lower MAE than each baseline across all jointly evaluated parameters, with Avg.\ NMAE of $0.078$ on single-material impacts and $0.116$ on blends. It runs in $6.1$ ms per query, compared with seconds for audio-language baselines and 190 ms for optimization-based estimation. These results demonstrate the accuracy and efficiency of our feed-forward approach under the evaluated settings.

\input{tables/ablation_table}
\input{tables/k_wise}
\vspace{-2mm}
\subsection{Objective and Multi-Impact Analysis}
\Cref{tab:ablation_checkmark} shows that no single auxiliary objective is sufficient for balanced continuous estimation. Direct regression gives the best single-material score, but its blend error remains higher, while adding the classification task does not improve the overall result. Using all objective terms balances single and blend performance, achieving the best blend and all-sample NMAE of $0.0989$ and $0.0950$, respectively. Adding the blend-related objectives reduces blend estimation error, while combining them with auxiliary classification yields the lowest blend and overall NMAE.

\Cref{tab:k_wise} further shows that using more reference impacts improves robustness to contact variation. As $K$ increases from 1 to 8, all-sample NMAE decreases from $0.1124$ to $0.0926$, with the same trend for both single and blend subsets. Since different hits vary in location, force, and timing, aggregating more recordings yields a more stable set embedding for the underlying material response.

\begin{figure}[t]
    \centering
    \includegraphics[width=1\linewidth]{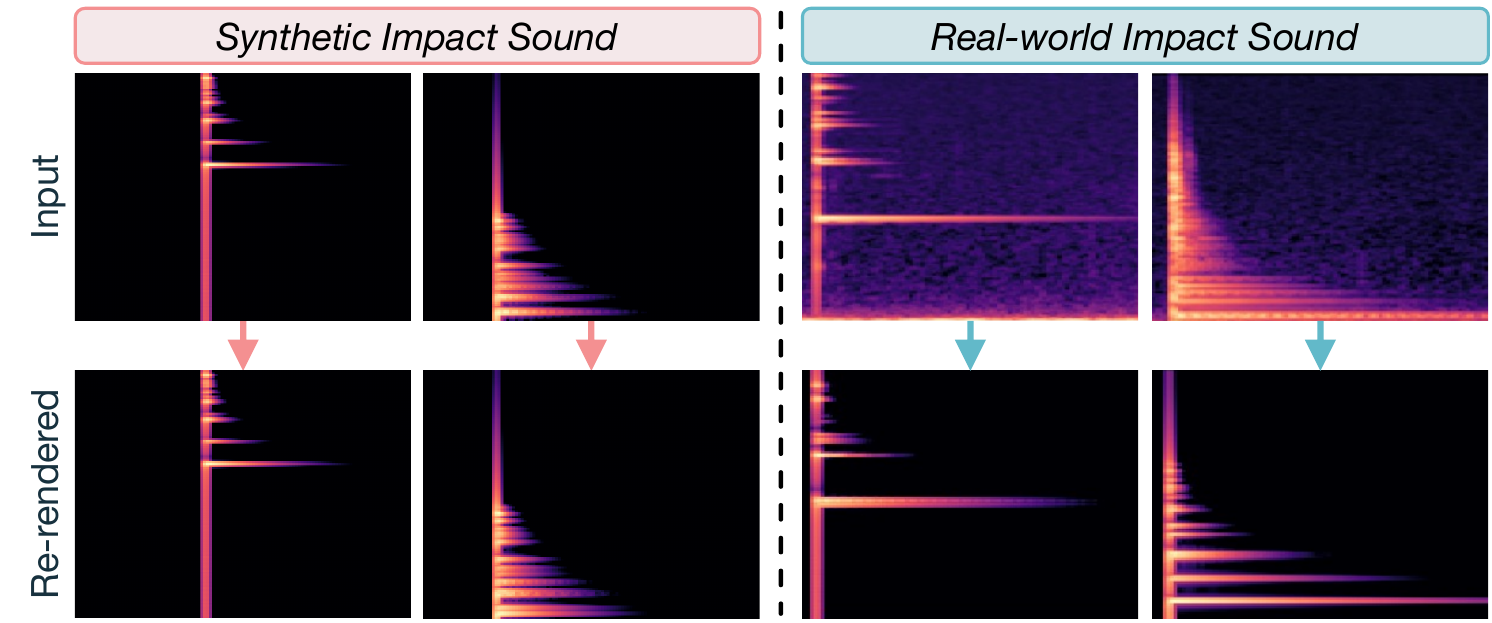}
    \caption{\textbf{Spectrogram comparison for re-rendering validation.} Each column compares an input reference with the sound synthesized after feeding the estimated parameters back into the simulator.}
    \label{fig:rerendering}
    \vspace{-4mm}
\end{figure}

\subsection{Real-world Re-rendering and Applications}
We evaluate our model on ObjectFolder-Real~\cite{gao2023objectfolder} to test whether predicted material parameters support re-rendering from real impact recordings. Because the dataset does not provide ground-truth material parameters, we evaluate the downstream rendered audio rather than direct parameter MAE. 
We combine the estimated material parameters with the corresponding object mesh provided by ObjectFolder-Real to synthesize impact sounds, and compare them with the reference recordings. 

We also assess perceptual material similarity through a listening study with 20 participants (\Cref{tab:userstudy}). Synthetic re-renderings are consistently judged similar to the input. On real recordings, our re-rendered sounds receive a 63.3\% similarity rate, comparable to the 59.0\% G.T. variant condition, where listeners compare two real recordings of the same object under different impacts. Since even real recordings of the same object can sound different under varying contact events, this result suggests that our re-renderings preserve plausible material cues under real-world variability.

\begin{figure}[t]
    \centering    \includegraphics[width=1\linewidth]{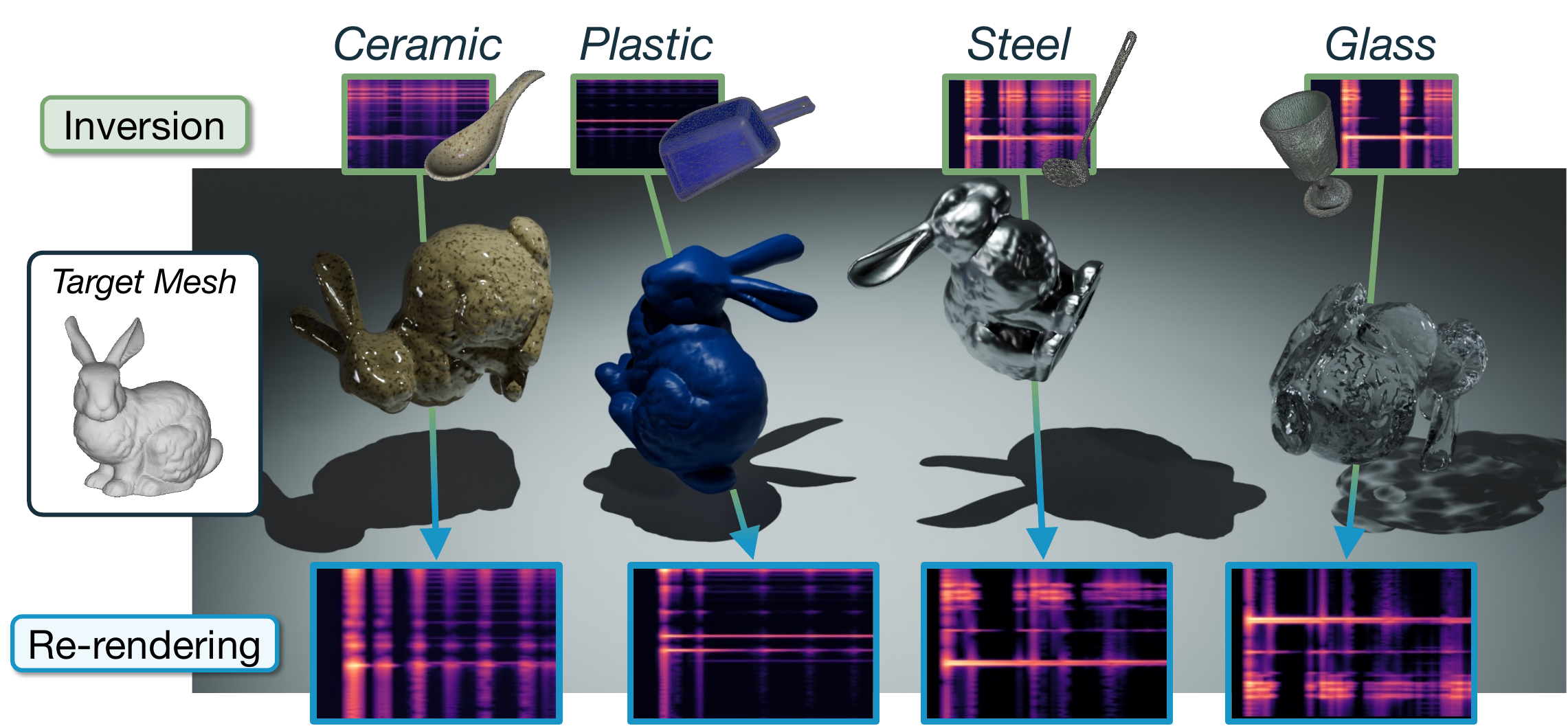}
    \caption{\textbf{Material transfer from reference audio to new geometry.} Estimated material parameters are transferred to the same target mesh to synthesize ceramic-, plastic-, steel-, and glass-like impact responses.}
    \label{fig:application}
    \vspace{-4mm}
\end{figure}

The same pipeline enables material transfer from reference audio. Given an impact sound, the model predicts renderer-compatible parameters that can be applied to a new target mesh. This supports VR/AR object interaction, game sound design, robotic manipulation, and digital-twin simulation, where objects must produce material-consistent impact sounds under different geometries and contact events. Fig.~\ref{fig:application} shows the same mesh re-rendered with ceramic-, plastic-, steel-, and glass-like acoustic characteristics.

\begin{table}[t]
    \centering
    \caption{\textbf{Listening study for perceptual material similarity.} Comparing reference and re-rendered impact sounds.}
    \label{tab:userstudy}
    \setlength{\tabcolsep}{4pt}
    \begin{tabular}{@{}llccc@{}}
    \toprule
    Domain & Condition & Sim.\% & Unsure & Diff. \\
    \midrule
    \multirow{1}{*}{Synthetic}
     & Re-render \textbf{(ours)} & 100\% & 0\% & 0\%  \\
    \midrule
    \multirow{2}{*}{Real}
     & G.T. variant & 59.0\%  & 17.0\% & 24.0\% \\
     & Re-render \textbf{(ours)} & 63.3\%  & 20.8\% & 15.8\%  \\
    \bottomrule
    \end{tabular}
\end{table}

%% file: tables/ablation_table.tex
\begin{table}[t]
\centering
\caption{\textbf{Ablation of training objectives for continuous estimation.} Checkmarks indicate the loss terms used during training; Single, Blend, and All report NMAE on the corresponding test subsets. Evaluated with \(K=4\), lower is better.}
\label{tab:ablation_checkmark}
\vspace{-4pt}
\setlength{\tabcolsep}{3pt}
\begin{tabular}{lcccccc}
\toprule
Variant & Reg. & Cls. & Blend  & Single & Blend & All \\
\midrule
- & $\checkmark$ &  &  & \textbf{0.0520} & 0.1053 & 0.0999 \\
- & $\checkmark$ & $\checkmark$ &  & 0.0648 & 0.1045 & 0.1005 \\
- & $\checkmark$ &  & $\checkmark$ &  0.0527 & 0.1032 & 0.0981 \\
\rowcolor{aliceblue}
Ours & $\checkmark$ & $\checkmark$ & $\checkmark$ & 0.0600 & \textbf{0.0989} & \textbf{0.0950} \\
\bottomrule
\vspace{-7mm}
\end{tabular}
\end{table}

%% file: tables/k_wise.tex
\begin{table}[t]
\centering
\caption{\textbf{Effect of multi-impact aggregation at inference.} Aggregating multiple reference impacts improves robustness to contact variation. Values report NMAE over the five material parameters. Lower is better.}
\label{tab:k_wise}
\vspace{-4pt}
\setlength{\tabcolsep}{16pt}
\begin{tabular}{lccc}
\toprule
K & Single & Blend & All \\
\midrule
1 & 0.0784 & 0.1163 & 0.1124 \\
2 & 0.0672 & 0.1043 & 0.1006 \\
4 & 0.0600 & 0.0989 & 0.0950 \\
\rowcolor{aliceblue} 8 & \textbf{0.0582} & \textbf{0.0965} & \textbf{0.0926} \\
\bottomrule
\vspace{-3mm}
\end{tabular}
\end{table}